\documentclass[twocolumn]{aastex631}

\shorttitle{A formation pathway of S0 galaxies}
\shortauthors{Zhou et al.}

\usepackage{graphics}

\usepackage{orcidlink}

\usepackage{etoolbox} % reference
\newrobustcmd{\disambiguate}[3]{#2 #3}

\usepackage{booktabs} % \bottomrule, \toprule

\usepackage{threeparttable}

\usepackage{scrextend}
\deffootnote[0.1em]{0em}{0.3em}{\thefootnotemark\space}

\edef\restoreparindent{\parindent=\the\parindent\relax}
\usepackage{parskip}
\restoreparindent

\begin{document}

\title{Misaligned gas acquisition as a formation pathway of S0 galaxies}

\author{Yuren Zhou~\orcidlink{0000-0001-7785-0626}}
\affiliation{School of Astronomy and Space Science, Nanjing University, Nanjing 210023, China}
\affiliation{Key Laboratory of Modern Astronomy and Astrophysics (Nanjing University), Ministry of Education, Nanjing 210023, China}
\affiliation{Collaborative Innovation Center of Modern Astronomy and Space Exploration, Nanjing 210023, China}

\author{Yanmei Chen~\orcidlink{0000-0003-3226-031X}}
\affiliation{School of Astronomy and Space Science, Nanjing University, Nanjing 210023, China}
\affiliation{Key Laboratory of Modern Astronomy and Astrophysics (Nanjing University), Ministry of Education, Nanjing 210023, China}
\affiliation{Collaborative Innovation Center of Modern Astronomy and Space Exploration, Nanjing 210023, China}

\author{Yong Shi~\orcidlink{0000-0002-8614-6275}}
\affiliation{School of Astronomy and Space Science, Nanjing University, Nanjing 210023, China}
\affiliation{Key Laboratory of Modern Astronomy and Astrophysics (Nanjing University), Ministry of Education, Nanjing 210023, China}
\affiliation{Collaborative Innovation Center of Modern Astronomy and Space Exploration, Nanjing 210023, China}

\author{Qiusheng Gu~\orcidlink{0000-0002-3890-3729}}
\affiliation{School of Astronomy and Space Science, Nanjing University, Nanjing 210023, China}
\affiliation{Key Laboratory of Modern Astronomy and Astrophysics (Nanjing University), Ministry of Education, Nanjing 210023, China}
\affiliation{Collaborative Innovation Center of Modern Astronomy and Space Exploration, Nanjing 210023, China}

\author{Junfeng Wang~\orcidlink{0000-0003-4874-0369}}
\affiliation{Department of Astronomy and Institute of Theoretical Physics and Astrophysics, Xiamen University, Xiamen, Fujian 361005, China}

\author{Dmitry Bizyaev~\orcidlink{0000-0002-3601-133X}}
\affiliation{Apache Point Observatory and New Mexico State University, PO Box 59, Sunspot, NM 88349-0059, USA}
\affiliation{Sternberg Astronomical Institute, Moscow State University, Moscow 119992, Russia}

\correspondingauthor{Yanmei Chen}
\email{chenym@nju.edu.cn}

\begin{abstract}
We analyze a sample of 753 S0 galaxies from the MPL-10 of MaNGA survey and investigate the gas-star kinematic misalignment and merger remnant fraction in galaxies with different morphological types. The misalign fraction in S0s is the highest among all the morphological types for both young (global $\mathrm{D}_n4000<1.6$, $\sim$15\%) and old (global $\mathrm{D}_n4000>1.6$, $\sim$10\%) galaxies. We compare the properties of misaligned S0s with other types of galaxies, finding: (i) misaligned S0s and misaligned spirals have higher bulge luminosity, higher B/T and larger S\'ersic index compared to spirals; (ii) the misaligned S0s have lower bulge luminosity $M_r$ and smaller bulge size than merger remnant S0s, while aligned S0s have the widest coverage for these parameter distributions which are overlapped with both misaligned S0s and merger remnant S0s; (iii) misaligned S0s have lower stellar mass $M_*$ and more isolated environment than aligned S0s and merger remnant S0s; (iv) the young misaligned S0s have positive $\mathrm{D}_n4000$ radial gradient, while the aligned S0s and merger remnant S0s show negative $\mathrm{D}_n4000$ radial gradient. Combining all these observational results, we suggest misaligned gas acquisition as another efficient formation pathway for S0 galaxies. The redistribution of gas angular momentum during gas-gas collision between accreted and pre-existing gas leads to gas inflow and the growth of bulge component, meanwhile the lack of cold gas at the outskirts leads to fading of spiral arms.
\end{abstract}

\section{Introduction}
\label{sec:introduction}
In the theory of galaxy formation and evolution, the morphology of galaxy provides the first clue to their evolutionary status. It is suggested that disks of galaxies are formed by cooling of gas in the dark matter halos \citep{blumenthal_formation_1984, mo_formation_1998}, while the bulges of galaxies are formed from mergers \citep{khochfar_origin_2006}. S0 galaxies, which are characterized by featureless disks and clear bulge components \citep{hubble_extragalactic_1926}, remain the subject of numerous debates regarding their formation pathways.

Two formation pathways have been proposed for S0 galaxies. The first one is fading of spiral galaxies \citep{gunn_infall_1972, larson_evolution_1980, quilis_gone_2000}. Observations show that the fraction of S0s in the local universe ($z\lesssim0.06$) increases with their projected local galaxy number density \citep[morphology-density relation]{dressler_galaxy_1980}, which suggests that the environmental processes, like starvation \citep{larson_evolution_1980} or ram-pressure stripping \citep{gunn_infall_1972, quilis_gone_2000}, reduce the gaseous content in spiral galaxies and lead to the formation of S0s. The second one includes mergers and galaxy interactions \citep{icke_distant_1985, bekki_unequal_1998, bekki_transformation_2011, querejeta_formation_2015}. The S0s are at least luminous in $K$-band as spiral galaxies \citep{burstein_kband_2005} and display systematically higher bulge-to-disk ratio \citep{dressler_galaxy_1980}. These evidences indicate that mechanisms leading to mass inflow and growth of bulge, such as mergers \citep{bekki_unequal_1998, querejeta_formation_2015} or galaxy interactions \citep{icke_distant_1985, bekki_transformation_2011}, play a role in the formation of S0s. However, neither pathway alone can explain all the properties of S0s. On the one hand, \cite{laurikainen_photometric_2010} find that the bulges of S0s exhibit properties more similar to spiral galaxies than elliptical galaxies in the $M_K^0$-$r_{\mathrm{eff}}$ diagram ($K$-band absolute magnitude $M_K^0$ and effective radius $r_{\mathrm{eff}}$ of bulge) and in the photometric plane (relation between S\'{e}rsic index $n$, $r_{\mathrm{eff}}$ and the central surface brightness of the bulge), indicating that mergers are not able to explain the statistical behaviour of bulges. On the other hand, \cite{wilman_morphological_2009} discover that S0s are as common in groups as in clusters at intermediate redshift ($z\sim0.4$) and tend to reside in the outskirts of groups. They propose processes like mergers, galaxy harassment and tidal interactions, which are effective at these locations, should be the most possible mechanisms for the formation of S0s. 

Using a sample of S0s from the SAMI Galaxy Survey, \cite{deeley_sami_2020} suggest rotationally supported S0s with $v/\sigma$ above $0.5$ form from faded spirals while pressure-supported S0s with $v/\sigma$ below $0.5$ form from mergers. Based on a sample of nearby S0s with single fiber spectra for the central regions from the Sloan Digital Sky Survey (SDSS) Data Release 7, \cite{xiao_nuclear_2016} find that star-forming (SF) S0s have the lowest stellar masses compared to other S0s and tend to locate in the low density environment. \cite{xu_starforming_2022} use a spatially resolved galaxies sample from the Mapping Nearby Galaxies at Apache Point Observatory (MaNGA) survey to show that SF S0s have bulges with significantly smaller S\'{e}rsic index than a normal S0s sample closely matched in $M_*$ and redshift. Both \cite{xiao_nuclear_2016} and \cite{xu_starforming_2022} propose external gas accretion or minor mergers as the primary formation mechanism of SF S0s. \cite{rathore_starforming_2022} find that the SF S0s show a distinct size-mass relation compared to spirals, which indicates fading of spirals is unlikely the formation mechanism of SF S0s. From the IllustrisTNG simulation, \cite{deeley_two_2021} suggest that 37\% of S0s form through gas stripping and 57\% form through mergers.

Long-slit spectral observations have revealed that the phenomenon of gas-star counter-rotation is ubiquitous ($\sim$20\%) in a small sample size ($<$70) of S0s \citep{bertola_external_1992, kuijken_search_1996, kannappan_broad_2001}. \cite{katkov_kinematics_2015} use the long-slit spectral observation from the Southern African Large Telescope to analyze a sample of S0s in isolated environment, finding that $\sim$39\%(=7/18) of them display a visible counter-rotation. Meanwhile, \cite{davis_atlas_2011} find out that $36\pm5$ percent of fast-rotating early-type galaxies in the $\text{ATLAS}^{\text{3D}}$ display gas-star kinematic misalignment. However, there are very few statistical studies of gas-star misalign fraction in S0s based on the integral field unit (IFU) survey due to the limited sample size. Thanks to the large IFU sample ($\sim$10,000) of MaNGA, we are able to study the gas-star kinematic misalignment fraction in S0s with different stellar population for the first time. The data analysis is displayed in Section \ref{section:data}. The comparison of the properties for misaligned S0s to other types of galaxies are in Section \ref{section:result}. We propose external misaligned gas acquisition as an efficient formation mechanism of S0 galaxies with global $\mathrm{D}_n4000<1.6$ in Section~\ref{section:discussion} and present our conclusions in Section~\ref{section:conclusion}.

\section{data analysis}
\label{section:data}
\subsection{MaNGA Survey}
MaNGA is one of three programmes in the fourth-generation SDSS (SDSS-IV) which starts on July 2014 \citep{bundy_overview_2015, drory_manga_2015, yan_sdss_2016, blanton_sloan_2017}. The science goal of MaNGA is to better understand how galaxies evolve and what regulates the star-formation, and it provides the spatially resolved spectroscopy with an unprecedented sample of $\sim$10,000 nearby galaxies at a spatial resolution of $\sim$1$\,\mathrm{kpc}$ at median redshift $z\sim0.03$. The targets are divided into `primary' and `secondary' samples, where the radial coverage is out to $\sim1.5$ effective (half-light) radius $R_e$ for the primary sample and $\sim2.5R_e$ for the secondary sample \citep{yan_sdss_2016}. The Data Reduction Pipeline \citep[DRP]{law_data_2016} offers sky-subtracted and flux-calibrated 3D spectra of each galaxy, while the Data Analysis Pipeline \citep[DAP]{westfall_data_2019} analyzes DRP-processed spectra to measure the stellar kinematics, nebular emission-line kinematics, and spectral indices such as $\mathrm{D}_n4000$.

The global stellar mass ($M_*$) are obtained from the NASA-Sloan Atlas\footnote{\url{http://nsatlas.org/}} \citep[NSA;][]{blanton_improved_2011} which is estimated from the spectral energy distribution (SED) fitting using the $K$-correction code. The global $\mathrm{D}_n4000$ is obtained by stacking the spectra with a median spectral signal-to-noise ratio (S/N) per pixel greater than 2 across the entire MaNGA bundle.

\subsection{Photometric Parameters and Morphological Classification}
\label{section:morphology}
The MaNGA PyMorph catalogue provides photometric parameters\footnote{PyMorph photometric Value Added Catalogue: \url{https://www.sdss4.org/dr17/data_access/value-added-catalogs/?vac_id=manga-pymorph-dr17-photometric-catalog}} and morphological classification\footnote{MaNGA Deep Learning Morphological Value Added Catalogue: \url{https://www.sdss4.org/dr17/data_access/value-added-catalogs/?vac_id=manga-morphology-deep-learning-dr17-catalog}} \citep{dominguez_sdss_2022}. The PyMorph photometric catalogue has single S\'{e}rsic fitting for S\'ersic index and two components S\'{e}rsic$+$Exponential fitting for bulge-to-total light ratio (B/T), bulge effective radius $R_e$ and bulge apparent magnitude. We use the $r$-band photometric parameters in this work and convert the $r$-band apparent magnitude $m_r$ to absolute magnitude $M_r$ based on the NSA redshift. Using the deep learning method, PyMorph morphological catalogue provides the numerical Hubble stage T-Type \citep[$\mathrm{T}<0$ corresponding to early-type galaxies]{deVaucouleurs_qualitative_1977}, probabilities for being late-type galaxies ($P_{\text{LTG}}$) and S0s ($P_{\text{S0}}$) based on the analysis of SDSS DR7 image data. This catalogue also provides visual inspection results as visual class (VC=2 for S0s) and visual flag (VF=0 for reliable visual class). By cross-matching MPL-10 galaxies with the PyMorph catalogue, we obtain 9247 galaxies in total. We adopt the same selection criterion outlined in \cite{dominguez_sdss_2022} to select S0s: $P_{\text{LTG}}<0.5$, $\text{T-Type}<0$, $P_{\text{S0}}>0.5$, $\text{VC}=2$ and $\text{VF}=0$. This criterion generates a sample of 789 S0s. We then check the images from the Dark Energy Spectroscopic Instrument \citep[DESI]{dey_overview_2019} survey to remove cases of on-going mergers, leading to a final sample of 753 S0s. We apply the same criteria as in \cite{dominguez_sdss_2022} to select 4534 spirals ($P_{\text{LTG}}>0.5$, $\text{T-Type}>0$, $\text{VC}=3$ and $\text{VF}=0$) and 2117 ellipticals ($P_{\text{LTG}}<0.5$, $\text{T-Type}<0$, $P_{\text{S0}}<0.5$, $\text{VC}=1$ and $\text{VF}=0$). 1843 galaxies which do not satisfy the above criteria are classified as others including irregulars or spiral galaxies viewing edge-on.

\subsection{Galaxies with Kinematic Misalignment}
\label{section:misalign_sample}
In order to obtain robust measurements on gas kinematics, we select emission-line galaxies as those with the H$\alpha$ emission-line S/N greater than 3 for at least 10 percent spaxels within $\sim1.5R_e$. Out of 9247 MPL-10 galaxies in the PyMorph catalogue, 6938 emission-line galaxies are selected. To quantify the kinematic misalignment between gas and stars, the position angle (PA) of kinematic major axis is determined using PaFit package in Python \citep{krajnovic_kinemetry_2006}. Misaligned galaxies are then defined as $\Delta{PA}\equiv|{PA}_{\text{gas}}-{PA}_{\text{star}}|>30^{\circ}$, where ${PA}_{\text{gas}}$ and ${PA}_{\text{star}}$ are PA for gaseous and stellar components, respectively. We select 723 candidates of misaligned galaxies with robust PA measurements ${PA}_{\text{error}}\leq20^{\circ}$, where ${PA}_{\text{error}}$ represents $\pm1\sigma$ error of PAs. 456 of them are visually verified as our final sample of misaligned galaxies, where ongoing mergers or galaxies with irregular morphology are removed. For more details about the selection of misaligned galaxies, we refer readers to \cite{zhou_sdss_2022}. 

\begin{figure*}
    \includegraphics[width=2.0\columnwidth]{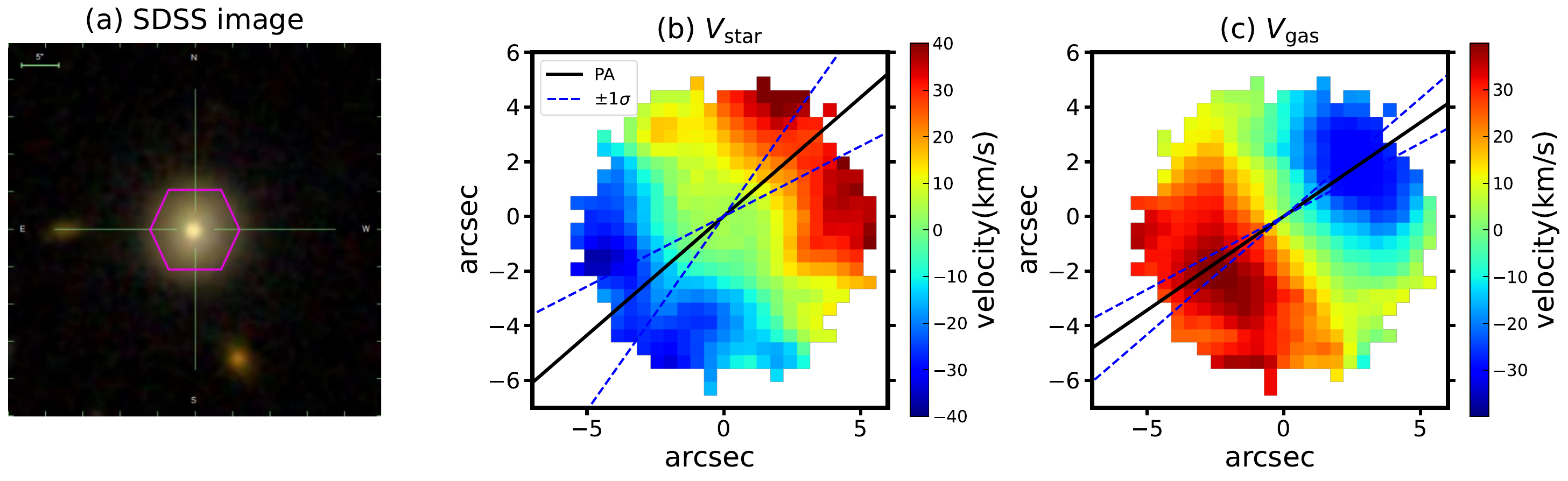}
    \caption{An example of misaligned S0 galaxy. Panel (a) displays the SDSS $g$, $r$, $i$-band false-color images with the purple hexagon marking the MaNGA bundle. Panel (b) and (c) show the velocity fields for the stellar and gaseous (traced by H$\alpha$) components, respectively. The red side moves away from us while the blue side moves towards us. The black solid lines mark the position angle of kinematic major axes with two blue dashed lines being their $\pm1\sigma$ errors.}
    \label{fig:misalign_sample}
\end{figure*}

Fig.~\ref{fig:misalign_sample} presents an example of misaligned S0 galaxy while panel (a) shows the SDSS false-color image and the purple hexagon marks the MaNGA bundle. Panels (b) and (c) show the stellar and gaseous velocity fields, respectively. The red side moves away from us while the blue side approaches us. The rotation or angular momentum direction of the gas and stars are clearly opposite. For each component, PA is marked by the black solid line while the two blue dashed lines indicate its $\pm1\sigma$ errors.

\subsection{Merger Remnant Features}
\label{section:merger_sample}
\begin{figure*}
    \includegraphics[width=2.0\columnwidth]{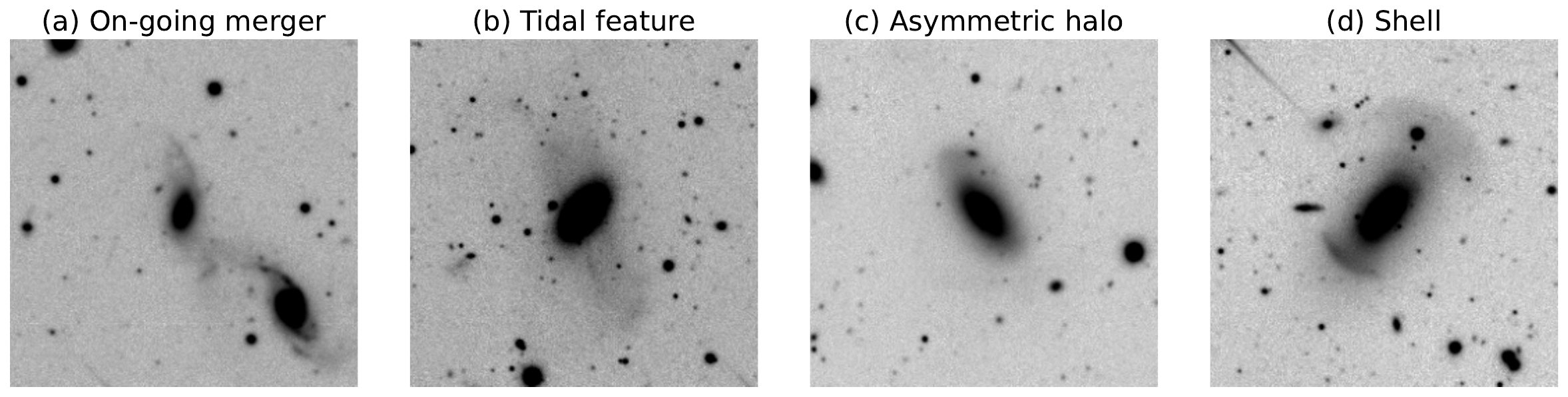}
    \caption{Classification of galaxies with interacting features: (a) on-going merger; (b) a galaxy with tidal features; (c) a galaxy with extended asymmetric stellar halo; (d) a galaxy with shell.}
    \label{fig:merger_sample}
\end{figure*}

Based on DESI survey, we search for faint merger remnant features of the 9247 MPL-10 galaxies in the PyMorph catalogue. The DESI images are $\sim$2$\,\mathrm{mag}$ deeper than SDSS in the $g$-, $r$-band. The DESI surveys cover $\sim$14000$\,\mathrm{deg}^2$ of sky and are composed of three public projects. These include the Beijing-Arizona Sky Survey \citep[BASS;][]{zou_project_2017}, which provides $g$-band and $r$-band images on the Bok 2.3-m telescope; the Mayall $z$-band Legacy Surveys \citep[MzLS;][]{silva_mayall_2016}, which provides $z$-band images on the 4-m Mayall telescope; the Dark Energy Camera Legacy Survey \citep[DECaLS;][]{blum_decam_2016}, which provides $g$-band, $r$-band and $z$-band images on the 4-m Blanco telescope. Following the procedure described in \cite{li_impact_2021}, we convolve the $r$-band image with a Gaussian kernel to match the resolution of the $g$-band image and stack them together to increase the S/N. Based on the stacked images, we divide galaxies into isolated and interacting ones. Fig.~\ref{fig:merger_sample} shows examples of galaxies with interacting features, including: (a) on-going mergers; (b) galaxies with tidal features; (c) galaxies with extended asymmetric stellar halos; (d) galaxies with shells. Our analysis identifies 737 galaxies with merger remnant features (b, c, d).

\subsection{Sample and Control Sample Selection}
\label{section:sample_selection}
\begin{table*}
\begin{center} 
  \caption{Classification of MPL-10 galaxies crossmatch with PyMorph catalogue in different morphological types}
  \begin{tabular}{c c c c c c}
  \toprule
  Type & total & spiral & S0  & elliptical  & other \\
  \hline
  MPL-10 galaxies in PyMorph & 9247 & 4534 & 753 & 2117 & 1843 \\
  Emission-line galaxies & 6938 & 4393 & 429 & 758 & 1358 \\
  Misaligned galaxies$^{\mathrm{(a)}}$ & 456 & 56 & 89 & 176 & 135 \\
  Galaxies with merger remnant features$^{\mathrm{(b)}}$ & 737 & 239 & 67 & 285 & 146 \\
  %Overlap between (a) \& (b) & 71 & 9 & 7 & 39 & 16\\
  \bottomrule
  \end{tabular}
  \label{tab:sample_selection}
\end{center}
\end{table*}

In this work, we focus on S0 galaxies and split them into three subgroups for comparison: (a) S0s with kinematic misalignment, which is the S0 sample in Section~\ref{section:morphology} crossmatch with the misaligned galaxy sample in Section~\ref{section:misalign_sample}; (b) S0s with merger remnant features, which is the S0 sample crossmatch with the merger remnant sample in Section~\ref{section:merger_sample}; (c) aligned S0s which are emission-line galaxies without kinematic misalignment. We check the potential overlap between galaxies in (a) and (b), finding that only 7 out of 753 S0s exhibit both misalignment and merger remnant features. Since merger can also cause kinematic misalignment, we put them into subgroup (b). The classification of MPL-10 galaxies with different morphology is listed in the Table~\ref{tab:sample_selection}.

In order to compare S0s with other morphological types, we build a non-S0 control sample with the minimal distance to the S0s in $\log M_*$ versus global $\mathrm{D}_n4000$ space, which is defined as $d=\sqrt{({\Delta\log M_*}/{0.1})^2+({\Delta\mathrm{D}_n4000}/{0.05})^2}$. $\Delta\log M_*$ and $\Delta\mathrm{D}_n4000$ are the difference in $\log M_*$ and $\mathrm{D}_n4000$ between a S0 galaxy and its control galaxy, respectively. We choose the global $\mathrm{D}_n4000$ instead of star-formation rate (SFR) to select the control sample because this parameter has a consistent measurement for all types of galaxies and avoids potential effect over the emission-line strength due to interactions between accreted and pre-existing gas in misaligned galaxies. Since the global $\mathrm{D}_n4000$ can reflect the age of galaxy stellar population, we split galaxies into young population with global $\mathrm{D}_n4000<1.6$ (young galaxies for short), as well as old population with global $\mathrm{D}_n4000>1.6$ (old galaxies).

\begin{figure*}
\begin{center}
    \includegraphics[width=1.2\columnwidth]{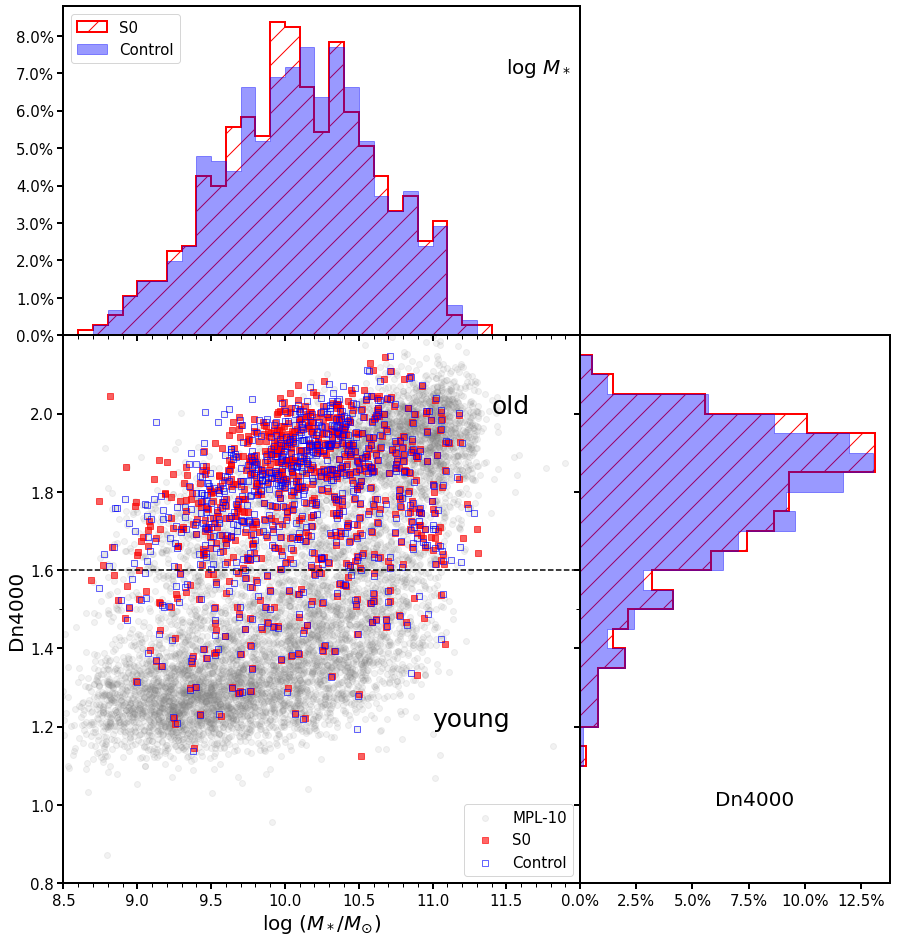}
    \caption{Global $\mathrm{D}_n4000$ versus $\log(M_*/M_{\odot})$ diagram for S0 galaxies and their non-S0 control sample, with the grey circles being MPL-10 galaxies. Red squares represent S0 galaxies and blue open squares are their corresponding controls. The horizontal lines correspond to $\mathrm{D}_n4000=1.6$ which split galaxies into young and old population. The top and right panels show distribution of stellar masses and global $\mathrm{D}_n4000$ for the S0 galaxies (red) and their control samples (blue) respectively, with each distribution is normalized with respect to their sample size.}
    \label{fig:control_sample}
\end{center}
\end{figure*}

Fig.~\ref{fig:control_sample} displays the global $\mathrm{D}_n4000$ versus $\log(M_*/M_{\odot})$ diagram for S0 galaxies (red squares) and their non-S0 control sample (blue open squares), with grey circles being MPL-10 galaxies. The horizontal dashed line corresponds to $\mathrm{D}_n4000=1.6$ which roughly splits galaxies into young and old population. The top and right panels show the distribution of stellar masses and global $\mathrm{D}_n4000$ for the S0s (red) and their controls (blue) respectively, with each distribution normalized to their sample size. 

\section{Results}
\label{section:result}
\subsection{Misalign and Merger Remnant Fraction}
\label{subsection:misalign_fraction}
\begin{figure*}
\begin{center}
    \includegraphics[width=2.0\columnwidth]{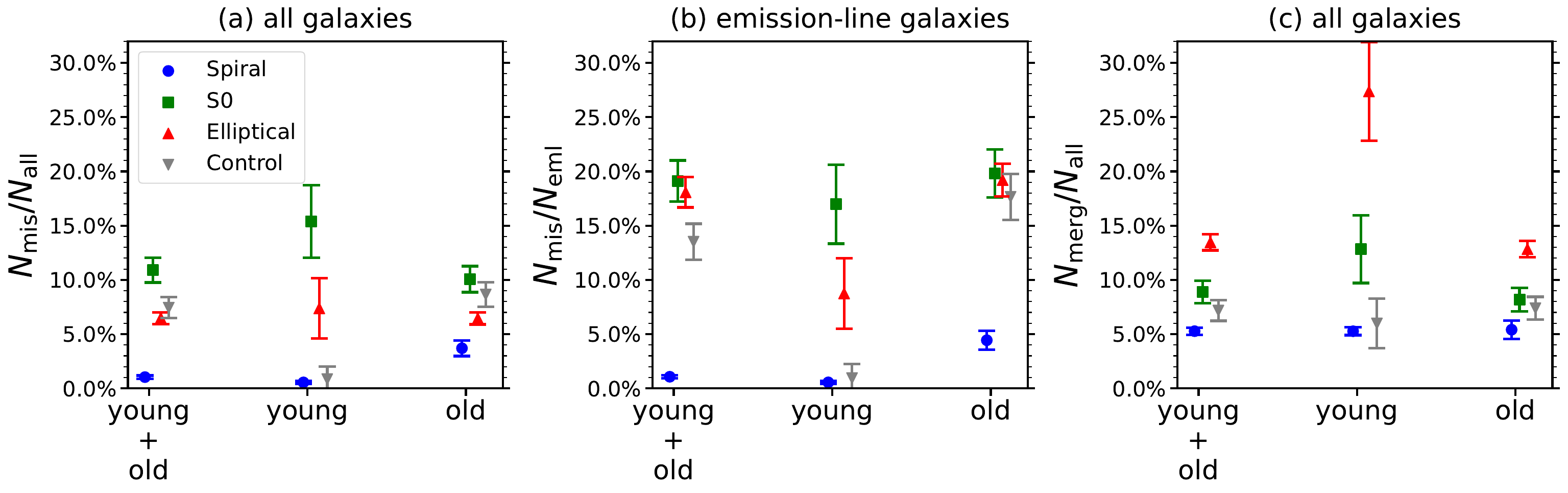}
    \caption{Panel (a) and (c) show the misalign fraction and merger remnant fraction in the entire MaNGA sample, respectively. Panel (b) shows the misalign fraction in the emission-line galaxies sample. Blue circles are for spiral galaxies, green squares for S0 galaxies and red triangles for elliptical galaxies. Each morphological type is divided into young and old population. Grey triangles represent the non-S0 control samples closely matched in $M_*$ and global $\mathrm{D}_n4000$ for S0 galaxies. Each point in the same population (young, old and young+old) is slightly offset in the $x$-axis to avoid overlap.}
    \label{fig:frac_mismerg}
\end{center}
\end{figure*}

We present a comparison of the gas-star misalign fraction (Fig.~\ref{fig:frac_mismerg}a \& \ref{fig:frac_mismerg}b) and merger remnant fraction (Fig.~\ref{fig:frac_mismerg}c) in galaxies which have different morphological types, including spirals (blue circles), S0s (green squares) and ellipticals (red triangles). To compare galaxies with different stellar population, we split galaxies into young and old population. We also include the non-S0 control sample built in Section~\ref{section:sample_selection} (grey triangles) in this comparison. In Fig.~\ref{fig:frac_mismerg}a, we show the misalign fraction which is defined as $N_{\mathrm{mis}}/N_{\mathrm{all}}$, where $N_{\mathrm{mis}}$ is the number of misaligned galaxies and $N_{\mathrm{all}}$ is the total number of galaxies (including galaxies with and without emission-lines) within a certain morphological type and stellar population bin. Since the kinematic misalignment can only be measured for emission-line galaxies due to the requirement of robust gas kinematics, we also present the comparison of misalign fraction $N_{\mathrm{mis}}/N_{\mathrm{eml}}$ in Fig.~\ref{fig:frac_mismerg}b, where $N_{\mathrm{eml}}$ is the number of emission-line galaxies within a certain morphological type and stellar population bin. The emission-line galaxies sample is only used in this section for comparison of misalign fraction and samples in other sections all contain lineless galaxies.

Fig.~\ref{fig:frac_mismerg}a shows that the fraction of S0s in general (young+old) population is $\sim$11\%, which is the highest compared to spirals ($\sim$1\%), ellipticals ($\sim$6\%) as well as the control sample ($\sim$8\%). Once we split the galaxies into young and old population, the differences between misalign fraction in S0s and other morphological types become more obvious for the young population ($\sim$15\% for the misaligned S0s), and less obvious for the old population ($\sim$10\% for the misaligned S0s). In Fig.~\ref{fig:frac_mismerg}b, the misalign fraction in emission-line S0s among the general (young+old) population is $\sim$19\%. This fraction is consistent with previous work using a small sample ($<70$) of S0s \citep[$\sim$20\%,][]{bertola_external_1992, kuijken_search_1996, kannappan_broad_2001} where kinematic misalignment is defined within emission-line galaxies. The misalign fraction is much higher in Fig.~\ref{fig:frac_mismerg}b than Fig.~\ref{fig:frac_mismerg}a for S0s and ellipticals with old stellar population. This is because nearly half of early-type galaxies (ETGs) are lineless. In summary, both Fig.~\ref{fig:frac_mismerg}a and Fig.~\ref{fig:frac_mismerg}b show that the young S0s have the highest misalign fraction, indicating a possible connection between kinematic misalignment and the formation of young S0s.

In Fig.~\ref{fig:frac_mismerg}c, we present merger remnant fraction in galaxies with different morphological types and stellar population. The merger remnant fraction in S0s is $\sim$9\% for the general population (young+old), lower than the $\sim$13\% for ellipticals, similar to that of spirals and non-S0 control sample. These trends are similar if we split the sample into young and old population. The difference of the merger remnant fraction in S0s and ellipticals becomes more obvious for young population. Comparing Fig.~\ref{fig:frac_mismerg}c \& \ref{fig:frac_mismerg}a, the misalign fraction in S0s is comparable to merger remnant fraction in S0s for the general population ($\sim$11\% vs. $\sim$9\%), and this is also true for young population ($\sim$15\% vs. $\sim$13\%) and old population ($\sim$10\% vs. $\sim$8\%). Considering that kinematic misalignment phenomenon is believed to be primarily contributed by external gas acquisition, this result may indicate that external gas acquisition in S0s is as equally efficient as mergers. We will discuss misaligned gas acquisition as a formation mechanism of S0s in Section~\ref{section:discussion}.

\subsection{Photometric Properties}
\begin{figure*}
\begin{center}
    \includegraphics[width=2.0\columnwidth]{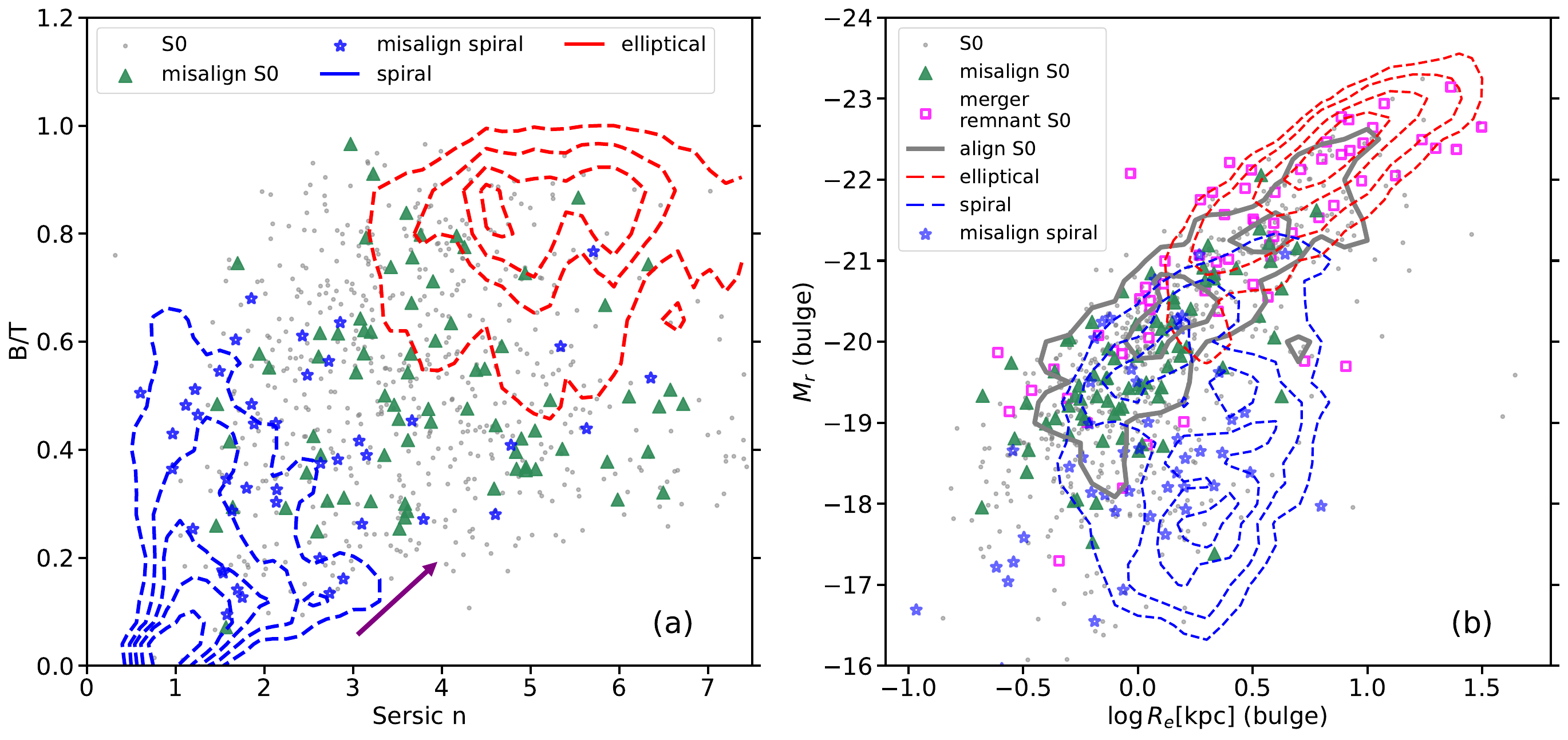}
    \caption{The panel (a) shows B/T vs. S\'ersic index n distribution. Grey dots represent the entire S0 sample, while the blue contours represent the spirals and red contours represent the ellipticals. The green triangles and blue stars are misaligned S0s and misaligned spirals, respectively. The purple arrow indicates the direction of a continuous distribution for misaligned spirals and misaligned S0s. The panel (b) shows the $r$-band bulge absolute magnitude $M_r$ vs. bulge effective radius $R_e$. Grey dots represent the entire S0 sample, while green triangles, magenta squares and blue stars are misaligned S0s, merger remnant S0s and misaligned spirals, respectively. The aligned S0s are displayed as grey solid contours while spiral galaxies and elliptical galaxies are shown as blue dashed and red dashed contours, respectively.}
    \label{fig:photometric_parameter}
\end{center}
\end{figure*}

To explore the impact of external gas acquisition on galaxy morphology, we compare the photometric properties of galaxies with different morphological types. Fig.~\ref{fig:photometric_parameter}a shows B/T vs. S\'ersic index $n$ distribution, grey dots represent the entire S0 sample, while the blue contours represent the spirals and red contours represent the ellipticals. The green triangles and blue stars are misaligned S0s and misaligned spirals, respectively. As we expect, misaligned S0s locate in between spirals and ellipticals, while they have significant higher S\'ersic index than misaligned spirals. We also find that misaligned spirals have higher B/T and S\'ersic index than the entire spiral sample, which indicates misaligned gas acquisition can lead to bulge growth.

Fig.~\ref{fig:photometric_parameter}b presents the $r$-band bulge absolute magnitude $M_r$ vs. bulge effective radius $R_e$. Grey dots represent the entire S0 sample, while green triangles, magenta squares and blue stars are misaligned S0s, merger remnant S0s and misaligned spirals, respectively. The aligned S0s are displayed as grey solid contours, while the entire spiral and elliptical samples are shown as blue dashed and red dashed contours, respectively. The misaligned S0s, aligned S0s and merger remnant S0s form a continuous sequence in bulge $M_r$ vs. $R_e$ diagram. Misaligned S0s dominate at lower bulge luminosity end with $\sim1.5$mag fainter compared to higher luminosity end in merger remnant S0s and they display significantly smaller bulge size, indicating misaligned S0s have different formation mechanism as merger. The aligned S0s have a wide coverage in bulge $M_r$ vs. $R_e$ diagram which are partially overlapped with both misaligned S0s and merger remnant S0s. Misaligned S0s have similar range of bulge $R_e$ as spirals and misaligned spirals, but at a certain radius they have highest bulge luminosity compared to lower value in misaligned spirals and the lowest value in the entire spiral sample, indicating the connection between kinematic misalignment and growth of bulge.

\subsection{Stellar and Dark Matter Halo Mass Distributions}
\begin{figure}
\begin{center}
    \includegraphics[width=1.0\columnwidth]{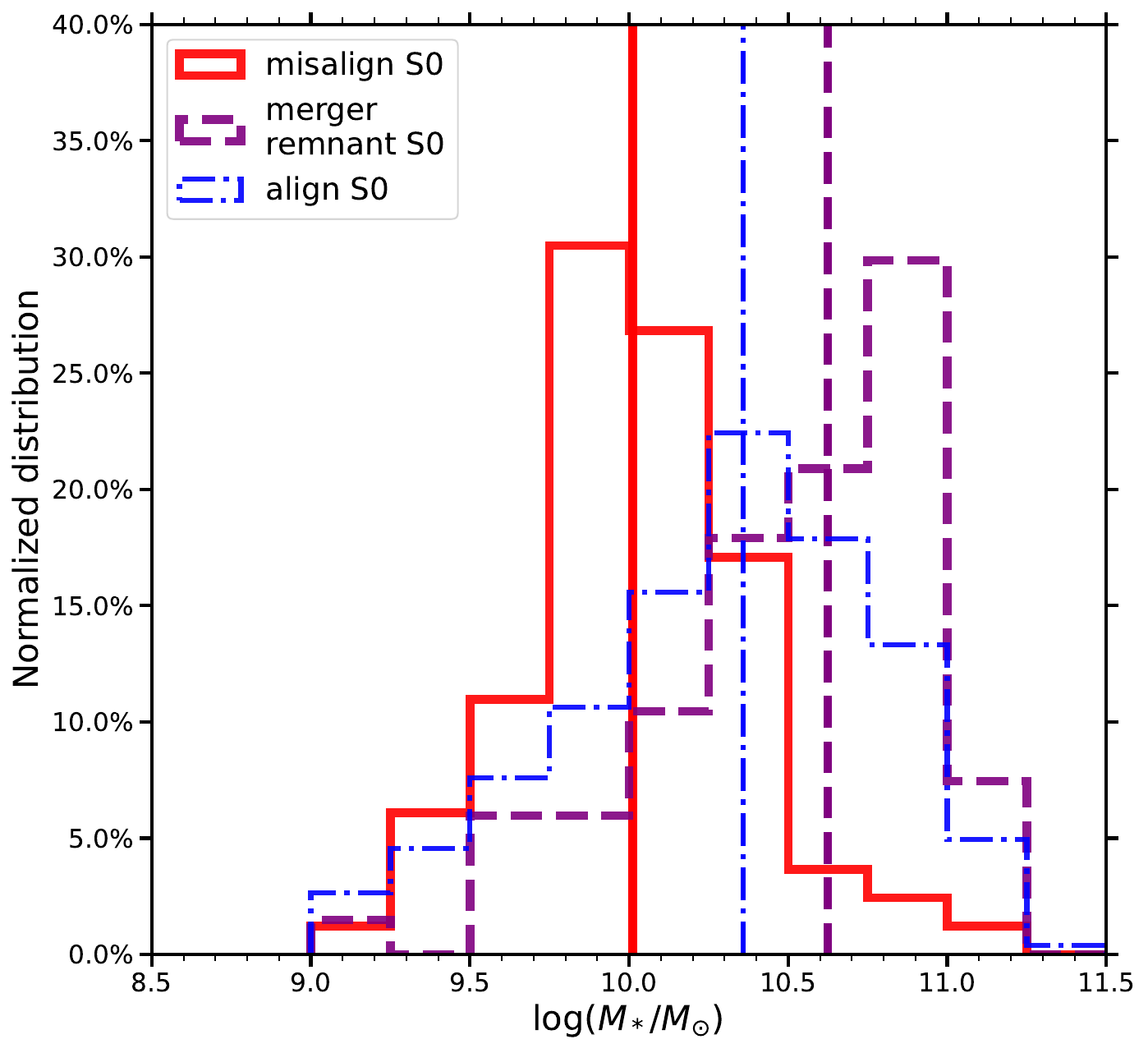}
    \caption{The distribution of stellar mass $M_*$ for misaligned S0s (red solid), aligned S0s (blue dashed-dotted) and merger remnant S0s (purple dashed). Each distribution is normalized to their sample size. The vertical lines with the same color as histograms are the median of the distributions.}
    \label{fig:mismerg_stellar_mass}
\end{center}
\end{figure}

We show the distribution of stellar mass $M_*$ in Fig.~\ref{fig:mismerg_stellar_mass} for misaligned S0s (red solid), aligned S0s (blue dashed-dotted) and merger remnant S0s (purple dashed). The vertical lines with the same color as the histograms represent the median of each distribution. Misaligned S0s have lower $M_*$ (median value of $\sim10^{10.0}M_{\odot}$) than merger remnant S0s (median value of $\sim10^{10.6}M_{\odot}$), while aligned S0s have widest $M_*$ distribution with a median value of $\sim10^{10.4}M_{\odot}$.

\begin{figure*}
\begin{center}
    \includegraphics[width=2.0\columnwidth]{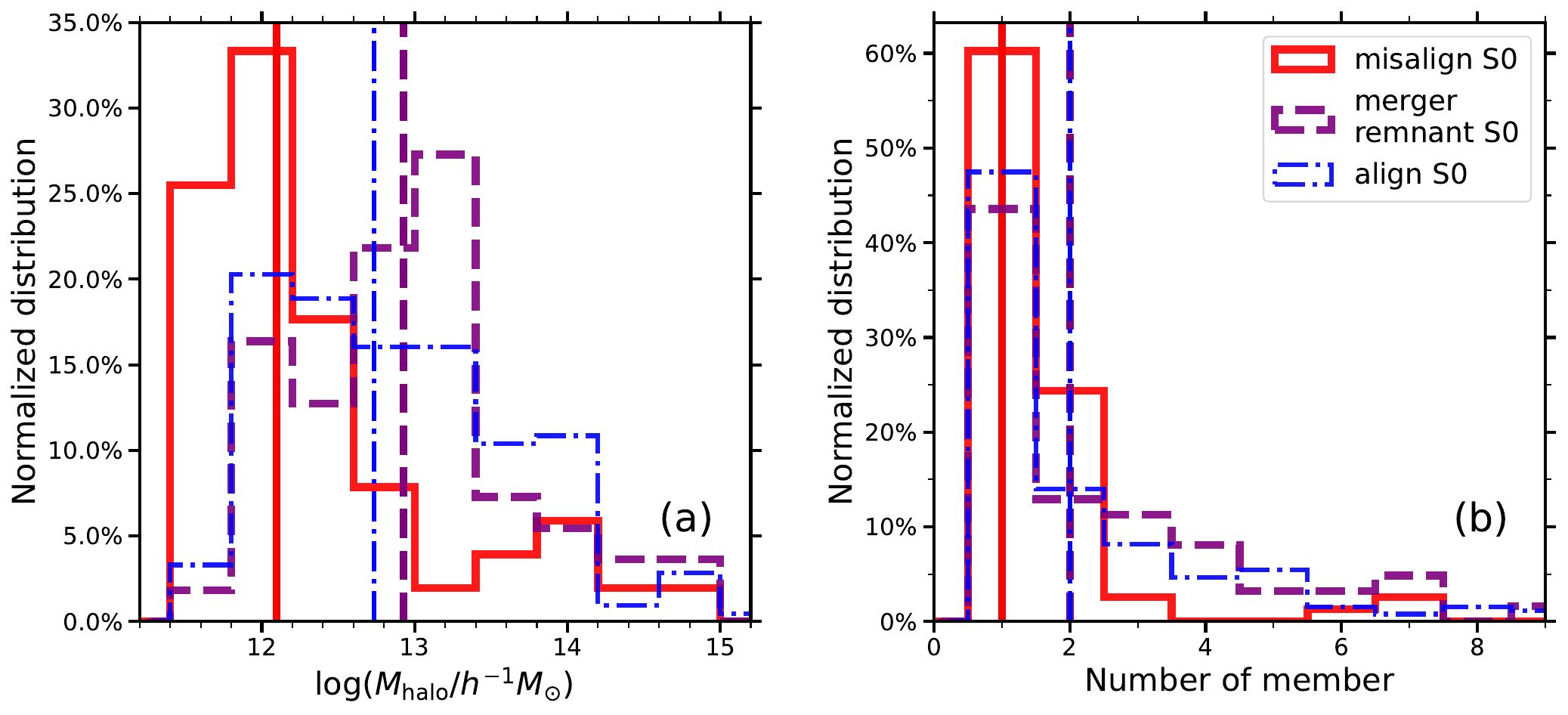}
    \caption{Panel (a) and (b) show the distribution of group halo mass $M_{\mathrm{halo}}$ and number of group members for galaxies within their galaxy group, respectively. Misaligned S0s, aligned S0s and merger remnant S0s are represented by red solid, blue dashed-dotted and purple dashed histograms, respectively. Each distribution is normalized to their sample size. The vertical lines with the same color as histograms are the median of the distributions.}
    \label{fig:mismerg_mass_halo}
\end{center}
\end{figure*}

We then compare the environmental properties by using group halo mass and the number of group members taken from \cite{yang_galaxy_2007}. They develop a method to categorize member galaxies into a galaxy group. Unlike the traditional friend-of-friend method, this technique can identify groups with only one member galaxy. They measure the group characteristic luminosity $L_{19.5}$, defined as the total luminosity of all group members down to $M_r-5\log h\le-19.5$ ($h=0.73$) with completeness correction. To obtain the halo mass of a given group, they first assume a one-to-one relation between the $L_{19.5}$ to halo masses. When a group halo mass function is given in a comoving volume, galaxy groups can be assigned halo masses according to their rank orders of $L_{19.5}$ (the highest halo mass is assigned to the galaxy group with the highest $L_{19.5}$). We present the distribution of group halo mass $M_{\mathrm{halo}}$ and the number of group members for each galaxy of its galaxy group in Fig.~\ref{fig:mismerg_mass_halo}a \& \ref{fig:mismerg_mass_halo}b, respectively. The misaligned S0s (red) tend to reside in lower mass host group halos (median value of $M_{\text{halo}}\sim10^{12}M_{\odot}$) and the merger remnant S0s (purple) tend to reside in higher mass host group halos (median value of $M_{\text{halo}}\sim10^{12.9}M_{\odot}$), while the aligned S0s (blue) cover a wider range of halo mass with a median value of $\sim10^{12.7}M_{\odot}$. In Fig.~\ref{fig:mismerg_mass_halo}b, misaligned S0s show the smallest number of member galaxies compared to merger remnant S0s and aligned S0s. Our results are consistent with \cite{davis_atlas_2011}, who discover that galaxies in dense groups and Virgo cluster always have aligned kinematics while $42\pm5$\% of fast-rotating ETGs have kinematic misalignment. Also, \cite{lagos_origin_2014} use the galaxy formation model GALFORM to show that gas-rich early-type galaxies tend to locate in low-mass halos.

\subsection{Stellar Population}
\label{subsection:ssp}
\begin{figure*}
\begin{center}
    \includegraphics[width=2.0\columnwidth]{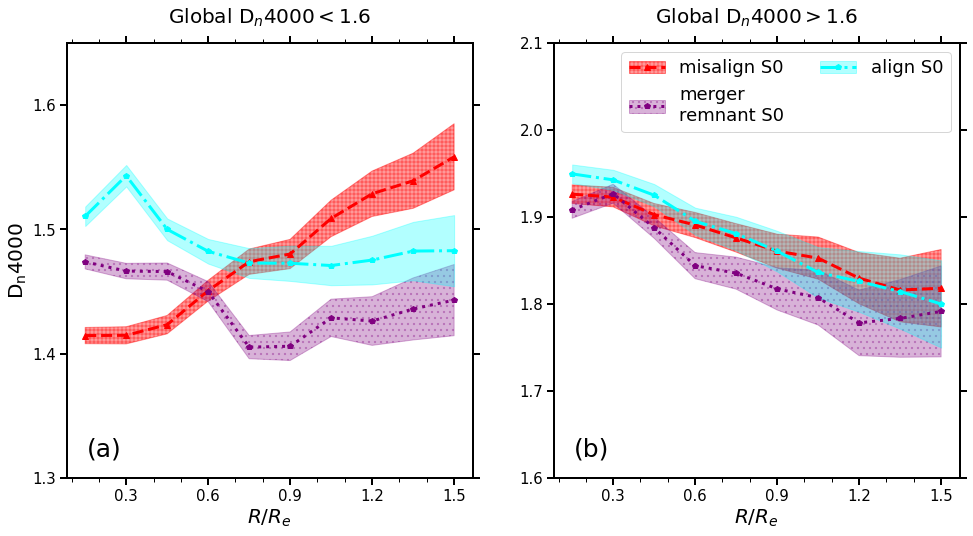}
    \caption{The $\mathrm{D}_n4000$ radial gradients for young (left) and old (right) misaligned S0s (red dashed), aligned S0s (cyan dashed-dotted) and merger remnant S0s (purple dotted). Thick lines represent the median of $\mathrm{D}_n4000$ in each $R/R_e$ bin while shaded regions with same color are the $\pm1\sigma$ error of $\mathrm{D}_n4000$ measurement.}
    \label{fig:Dn4000_median_expcont}
\end{center}
\end{figure*}

Fig.~\ref{fig:Dn4000_median_expcont} presents the radial gradient of $\mathrm{D}_n4000$ for S0s with different types. Thick lines represent the median of $\mathrm{D}_n4000$ in each $R/R_e$ bin while shaded regions with the same color give the $\pm1\sigma$ error of $\mathrm{D}_n4000$ measurement. Red dashed lines with hollowed squares are misaligned S0s, cyan dashed-dotted lines are aligned S0s while purple dotted lines with filled dots are merger remnant S0s. The left and right panel show galaxies with young and old stellar population, respectively.

Fig.~\ref{fig:Dn4000_median_expcont}a shows the young misaligned S0s have positive $\mathrm{D}_n4000$ radial gradient, which indicates younger stellar population in the central region compared to their outskirts. The aligned S0s and merger remnant S0s show negative $\mathrm{D}_n4000$ radial gradient. The lower $\mathrm{D}n4000$ for merger remnant S0s than aligned S0s could be due to merger induced star-formation. In Fig.~\ref{fig:Dn4000_median_expcont}b, the old misaligned S0s, aligned S0s and S0s with merger remnant features display very similar $\mathrm{D}_n4000$ radial profile with a clear negative radial gradient.

\section{Discussion}
\label{section:discussion}
In Section~\ref{subsection:misalign_fraction}, we show that misalign fraction in S0s is the highest among all the morphological types. Therefore, this leads to the question whether there is any connection between kinematic misalign phenomena and the formation of S0s?

Based on the IllustrisTNG simulation, \cite{deeley_sami_2020} point out merger and gas stripping in dense environment (faded spiral) are two dominated mechanisms in the formation of S0s. However, neither of these two mechanisms can explain the properties of misaligned S0s. On the one hand, misaligned S0s have significantly lower bulge luminosity and smaller bulge size compared to merger remnant S0s or ellipticals (see Fig.~\ref{fig:photometric_parameter}), indicating major merger is not the dominated formation mechanism of misaligned S0s. On the other hand, comparing with aligned S0s and merger remnant S0s, misaligned S0s tend to locate in isolated environment (Fig.~\ref{fig:mismerg_mass_halo}), indicating gas stripping which tends to happen in dense environment is not the dominated formation mechanism of misaligned S0s. Meanwhile the positive $\mathrm{D}_n4000$ radial gradient of misaligned S0s (Fig.~\ref{fig:Dn4000_median_expcont}) is not able to be explained by faded spiral if gas is stripped at outskirts.

Putting all these observational results together with the truth that S0s have the highest misalign fraction among all the galaxy morphology type, we propose misaligned gas acquisition scenario for the formation of misaligned S0s, especially in young galaxies. The accreted gas interacts with pre-existing gas leading to angular momentum redistribution which triggers gas inflow and the central star-formation \citep{jin_sdss_2016, chen_growth_2016, xu_sdss_2022, zhou_sdss_2022}.

\subsection{Bulge Growth from Misaligned Gas Acquisition}
\label{section:bulge_growth}
\begin{figure*}
\begin{center}
    \includegraphics[width=1.5\columnwidth]{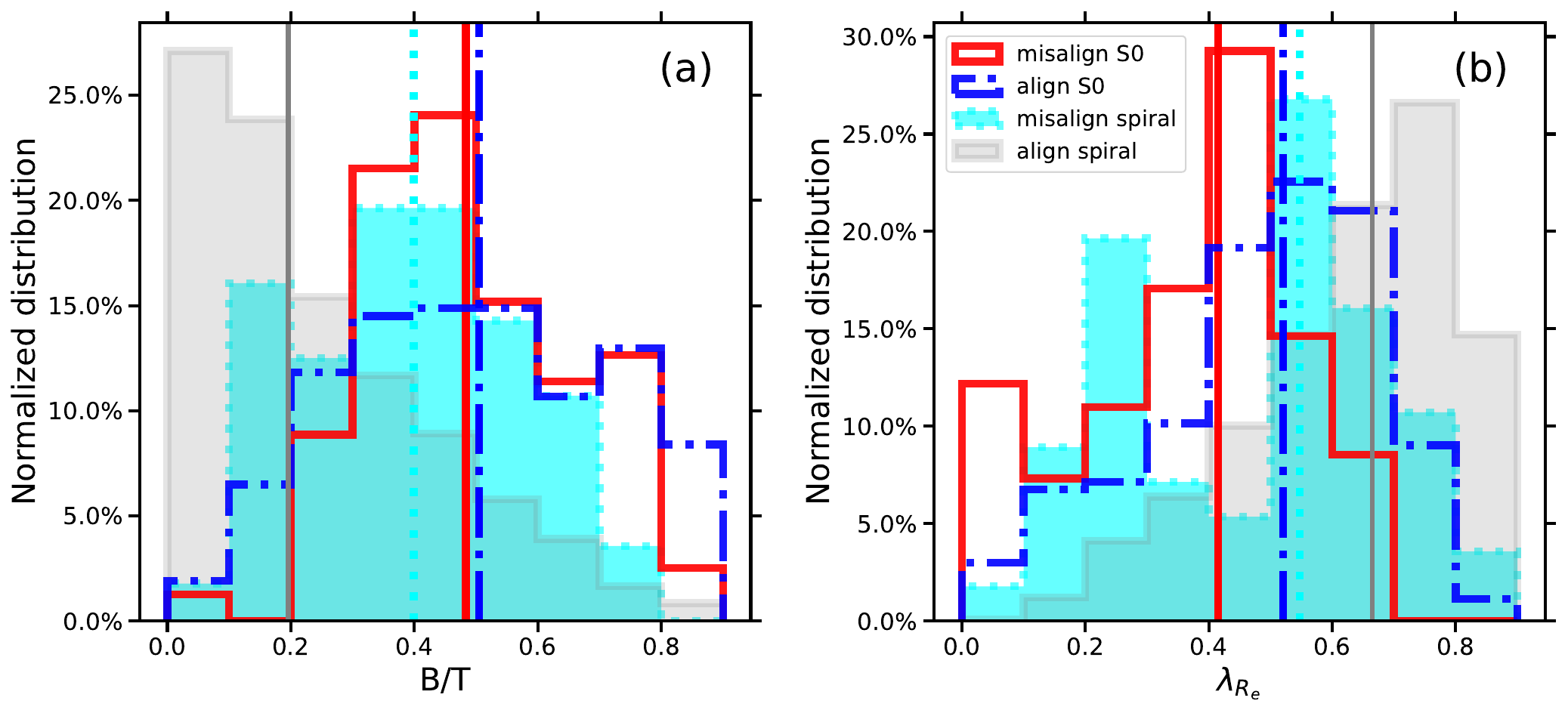}
    \caption{Panel (a) shows B/T distribution and panel (b) shows the $\lambda_{R_e}$ distribution. Misaligned S0s, aligned S0s, misaligned spirals and aligned spirals are represented by red solid, blue dashed-dotted, cyan shaded and grey shaded, respectively. Each distribution is normalized to their sample size. The vertical lines with the same color as histograms are the median of the distributions.}
    \label{fig:BtoT_lambda_Re}
\end{center}
\end{figure*}

To explore the influence of external gas acquisition on bulges growth, we compare the B/T (panel a) and $\lambda_{R_e}$ (panel b) in Fig.~\ref{fig:BtoT_lambda_Re} for misaligned S0s (red solid), aligned S0s (blue dashed-dotted), misaligned spirals (cyan shaded) and aligned spirals (grey shaded). The vertical lines with the same color as the histograms are the median of the distributions. The spin parameter $\lambda_{R_e}$ is a proxy of the specific angular momentum for stars defined as
\begin{equation}
\lambda_{R_e}=\frac{\sum^{N_p}_{i=1}F_iR_i|V_i|}{\sum^{N_p}_{i=1}F_iR_i\sqrt{V_i^2+\sigma_i^2}},
\end{equation}
where $F_i$, $V_i$, $\sigma_i$ are the $r$-band flux, stellar velocity and stellar velocity dispersion of the $i$-th spaxel within $R_e$. $R_i$ is the distance between the $i$-th spaxel and the galaxy center. Larger $\lambda_{R_e}$ means higher specific angular momentum \citep{emsellem_sauronIX_2007}. Fig.~\ref{fig:BtoT_lambda_Re}a shows that the misaligned spirals have similar B/T as S0s (misaligned and aligned), which is much higher than aligned spirals. Fig.~\ref{fig:BtoT_lambda_Re}b shows that both misaligned S0s and misaligned spirals have much lower $\lambda_{R_e}$ than aligned spirals, while their $\lambda_{R_e}$ is similar or even lower than aligned S0s. One natural explanation to the higher B/T and lower $\lambda_{R_e}$ in misaligned S0s and misaligned spirals in Fig.~\ref{fig:BtoT_lambda_Re} is misaligned gas acquisition, where the accreted gas interacts with pre-existing gas leading to gas inflow and the growth of bulge component.

\subsection{Lower Spiral Arm Fraction in Misaligned Galaxies}
\label{section:spiral_fading}
\begin{figure}
\begin{center}
    \includegraphics[width=1.0\columnwidth]{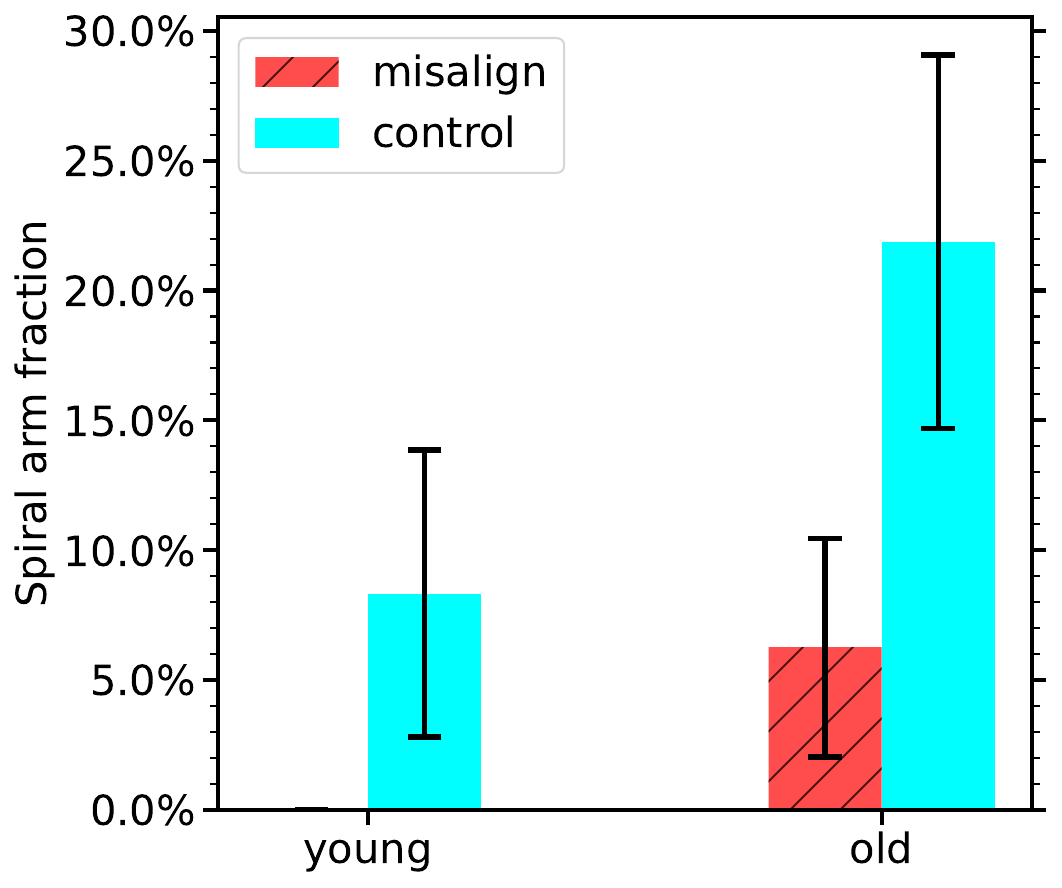}
    \caption{Fraction of galaxies with spiral arm features in misaligned and aligned control galaxies with different stellar population. Red histograms with hatched lines are misaligned galaxies while the cyan histograms are aligned controls.}
    \label{fig:fraction_spiral}
\end{center}
\end{figure}

The formation of S0s includes growth of bulge and fading of spiral arm features. In Section~\ref{section:bulge_growth}, we show the impact of misaligned gas acquisition on the growth of bulge. In this section, we explore the impact of misaligned gas acquisition on the spiral arm fading. Since PyMorph catalogue does not classify disk galaxies from spirals, we visually check the SDSS image to separate disk galaxies from spirals. For misaligned spirals, we build an control sample which is aligned spirals with similar $M_*$, global $\mathrm{D}_n4000$, inclination angle and redshift. Fig.~\ref{fig:fraction_spiral} shows the fraction of galaxies with spiral arm features in misaligned and aligned control samples. Red histograms with hatched lines are misaligned galaxies while the cyan histograms are aligned controls. For both young and old samples, the fraction of galaxies with spiral arm features is much lower in misaligned galaxies than that in aligned controls. It is intriguing that none of the young misaligned spirals (classified as spirals by PyMorph) shows spiral arm features. This suggests misaligned gas acquisition contributes to the formation of disk galaxies, where quenching happens at the outskirts of galaxies indicated by the positive $\mathrm{D}_n4000$ gradient of misaligned S0s in Fig.~\ref{fig:Dn4000_median_expcont}a.

Simulation works also shed light on the relation between spiral arms fading and gas-star counter-rotation. Based on the $N$-body simulations GADGET-3, \cite{donghia_self_2013} compare two galaxies with only difference in the rotational direction of gaseous components, finding that the spiral arms cannot form once the gaseous components are counter-rotating. \cite{osman_strong_2017} obtain similar results through smoothed-particle hydrodynamical simulation, suggesting that the existence of counter-rotating gas suppress the swing amplification mechanism for the formation of spiral arms~\citep{toomre_amplifies_1981, michikoshi_swing_2016}. These simulations are qualitatively consistent with our results.

Combining observational results in Section~\ref{section:bulge_growth} \& \ref{section:spiral_fading}, we suggest misaligned gas acquisition have significant impact over galaxy morphology with growth of bulge and fading of spiral arms, which leads to formation of S0s.

\subsection{The Contribution of Mergers in Misaligned Galaxies}
\label{subsection:timescale}
The visibility timescale of merger remnant features depends on both the observation depth and the time at which merging process happened. Using N-body hydrodynamic simulations GADGET2, \cite{ji_lifetime_2014} find that the visibility timescale of merger remnant features is $\gtrsim1.4~\text{Gyr}$ for $r$-band surface brightness limit $\mu_r=25~\text{mag}~\text{arcsec}^{-2}$, while the timescale is $\gtrsim5~\text{Gyr}$ for $\mu_r=28~\text{mag}~\text{arcsec}^{-2}$. We observe 66 misaligned galaxies with APO-3.5m telescope in the white-light mode, where the surface brightness limit is $\sim$28$~\text{mag}~\text{arcsec}^{-2}$, finding that $<5\%$ of these misaligned galaxies show merger remnant features. The surface brightness limit for the DESI survey, which is used for our merger remnant classification, is $\sim27~\text{mag}~\text{arcsec}^{-2}$. Therefore, combining this result with significant differences between bulge luminosity and radius between misaligned S0s and merger remnant S0s (see Fig.~\ref{fig:photometric_parameter}b), we conclude mergers are not the main driver in the formation of misaligned S0s.

Observational and simulation studies also reveal that although mergers give rise to kinematic misalignment, it is not a dominated mechanism. From observational studies, \cite{li_impact_2021} find that the merger remnant fraction are similar between SF misaligned galaxies and aligned galaxies, indicating mergers are not the primary origin of misalignment in SF galaxies. Meanwhile the merger remnant fraction in quiescent (QS) misaligned galaxies is $\sim$10\% higher than the merger remnant fraction in aligned galaxies, indicating mergers have contribution to misalignment in QS galaxies, but not the dominated mechanism. From simulation studies, \cite{lagos_origin_2015} use the GALFORM model of galaxy formation to investigate misalign fraction in ETGs, finding that the misalign fraction is 2--5\% with only merger process, when they add gas accretion, this fraction increases to $\sim$46\% which is consistent with $\mathrm{ATLAS}^{3D}$ observational result ($\sim$42\% for ETGs). Based on the Illustris simulations, \cite{starkenburg_origin_2019} find that mergers alone cannot account for the misaligned fraction in low-mass galaxies with $\log(M_*/M_{\odot})<10.7$. Therefore, we prefer gas accretion as the primary mechanism for the formation of misaligned galaxies.

\section{Conclusions}
\label{section:conclusion}
We analyze a sample of 753 S0 galaxies from the MPL-10 of MaNGA survey and investigate the gas-star kinematic misalignment fraction and merger remnant fraction in galaxies with different morphological types. We compare the $r$-band photometric parameters, stellar mass, group halo mass and number of group members as well as stellar population between misaligned S0s and other types of galaxies. The results are the following:

(i) the misalign fraction in S0s is the highest among all the morphological types for both young and old galaxies, and it is comparable to merger remnant fraction in S0s;

(ii) in the bulge $M_r$ vs. $R_e$ diagram, misaligned S0s dominate at lower bulge luminosity end with significantly smaller bulge size compared to higher luminosity end in merger remnant S0s. The aligned S0s have a wide coverage in $M_r$ vs. $R_e$ diagram which are partially overlapped with both misaligned S0s and merger remnant S0s.

(iii) misaligned S0s have lower $M_*$ (median value of $\sim10^{10.0}M_{\odot}$) than merger remnant S0s (median value of $\sim10^{10.6}M_{\odot}$), while aligned S0s have widest $M_*$ distribution with a median value of $\sim10^{10.4}M_{\odot}$;

(iv) misaligned S0s tend to reside in lower mass host group halos with lower number of member galaxies and merger remnant S0s tend to reside in higher mass host group halos with higher number of member galaxies, while the aligned S0s cover a wider range of halo mass and number of member galaxies;

(v) the young misaligned S0s have positive $\mathrm{D}_n4000$ radial gradients, indicating younger stellar population in the central region than the outskirts;

(vi) misaligned spirals have similar B/T as S0s (misaligned and aligned), which is much higher than aligned spirals, meanwhile both misaligned S0s and misaligned spirals have much lower $\lambda_{R_e}$ than aligned spirals and their $\lambda_{R_e}$ is similar or even lower than aligned S0s.

(vii) misaligned galaxies have lower fraction of spiral arm features than that of aligned galaxies;

Since misaligned S0s have distinct features compared to merger remnant S0s in bulge $M_r$ vs. $R_e$ diagram, $M_*$ distribution as well as group environment properties, we conclude misaligned S0s have different formation mechanism as merger remnant S0s. Meanwhile, misaligned S0s locate in isolated environment, faded spiral is not the dominated formation mechanism of misaligned S0s. We propose misaligned gas acquisition as an efficient formation pathway for young S0 galaxies. The accreted gas interacts with pre-existing gas in galaxies, leading to angular momentum redistribution and gas inflow, which triggers central star-formation and bulge growth (ii, v, vi). Meanwhile, the lack of cold gas at the outskirts leads to fading of spiral arms (vii).

\hspace*{\fill}
\section*{Acknowledgements}
    Y. C acknowledges support from the NSFC grant 12333002 and the China Manned Space Project (grant No. CMS-CSST-2021-A05) as well as the China Manned Space Project (the second-stage CSST science project: ``Investigation of small-scale structures in galaxies and forecasting of observations"). J.W. acknowledges support by the NSFC grants U1831205, 12033004 and 12221003. DB is partly supported by RSCF grant 22-12-00080.
    
This work is partly based on observations obtained with the Apache Point Observatory 3.5-meter telescope, which is owned and operated by the Astrophysical Research Consortium.

Funding for the Sloan Digital Sky 
Survey IV has been provided by the 
Alfred P. Sloan Foundation, the U.S. 
Department of Energy Office of 
Science, and the Participating 
Institutions. 

SDSS-IV acknowledges support and 
resources from the Center for High 
Performance Computing  at the 
University of Utah. The SDSS 
website is www.sdss.org.

SDSS-IV is managed by the 
Astrophysical Research Consortium 
for the Participating Institutions 
of the SDSS Collaboration including 
the Brazilian Participation Group, 
the Carnegie Institution for Science, 
Carnegie Mellon University, Center for 
Astrophysics | Harvard \& 
Smithsonian, the Chilean Participation 
Group, the French Participation Group, 
Instituto de Astrof\'isica de 
Canarias, The Johns Hopkins 
University, Kavli Institute for the 
Physics and Mathematics of the 
Universe (IPMU) / University of 
Tokyo, the Korean Participation Group, 
Lawrence Berkeley National Laboratory, 
Leibniz Institut f\"ur Astrophysik 
Potsdam (AIP),  Max-Planck-Institut 
f\"ur Astronomie (MPIA Heidelberg), 
Max-Planck-Institut f\"ur 
Astrophysik (MPA Garching), 
Max-Planck-Institut f\"ur 
Extraterrestrische Physik (MPE), 
National Astronomical Observatories of 
China, New Mexico State University, 
New York University, University of 
Notre Dame, Observat\'ario 
Nacional / MCTI, The Ohio State 
University, Pennsylvania State 
University, Shanghai 
Astronomical Observatory, United 
Kingdom Participation Group, 
Universidad Nacional Aut\'onoma 
de M\'exico, University of Arizona, 
University of Colorado Boulder, 
University of Oxford, University of 
Portsmouth, University of Utah, 
University of Virginia, University 
of Washington, University of 
Wisconsin, Vanderbilt University, 
and Yale University.

\bibliographystyle{aasjournal}
\bibliography{reference.bib}

\end{document}